\documentclass[aps,pra,floatfix,superscriptaddress,twocolumn,reprint,tigthenlines]{revtex4-1}

\usepackage{amsmath}
\usepackage{amssymb}
\usepackage{graphicx}
\usepackage{algorithm}
\usepackage{algorithmic}
\usepackage{units}
\usepackage{dcolumn}
\usepackage{bm}
\usepackage{epsfig}
\def\comment#1{}

\usepackage{color} 
\usepackage[normalem]{ulem}

\begin{document}

\title{Photonic Quantum Networks formed from NV$^-$ centers}

\author{Kae Nemoto}\email{nemoto@nii.ac.jp}
\affiliation{National Institute of Informatics, 2-1-2 Hitotsubashi, Chiyoda-ku, Tokyo 101-8430, Japan}

\author{Michael Trupke}
\affiliation{Vienna Center for Quantum Science and Technology, Atominstitut, TU Wien, 1020 Vienna, Austria}

\author{Simon J. Devitt}
\affiliation{National Institute of Informatics, 2-1-2 Hitotsubashi, Chiyoda-ku, Tokyo 101-8430, Japan}

\author{Burkhard Scharfenberger}
\affiliation{National Institute of Informatics, 2-1-2 Hitotsubashi, Chiyoda-ku, Tokyo 101-8430, Japan}

%\author{Ashley M. Stephens}
%\affiliation{National Institute of Informatics, 2-1-2 Hitotsubashi, Chiyoda-ku, Tokyo 101-8430, Japan}

\author{Kathrin Buczak}
\affiliation{Vienna Center for Quantum Science and Technology, Atominstitut, TU Wien, 1020 Vienna, Austria}

%\author{Tobias N\"obauer}
%\affiliation{Vienna Center for Quantum Science and Technology, Atominstitut, TU Wien, 1020 Vienna, Austria}

\author{J\"org Schmiedmayer}
\affiliation{Vienna Center for Quantum Science and Technology, Atominstitut, TU Wien, 1020 Vienna, Austria}

\author{William J. Munro}
\affiliation{NTT Basic Research Laboratories, NTT Corporation, 3-1 Morinosato-Wakamiya, Atsugi, Kanagawa 243-0198, Japan}
\affiliation{National Institute of Informatics, 2-1-2 Hitotsubashi, Chiyoda-ku, Tokyo 101-8430, Japan}

\begin{abstract} 
In this article we present a simple repeater scheme based on the negatively-charged nitrogen vacancy centre in diamond (NV$^-$). Each repeater node is built from simple modules comprising an optical cavity containing a single NV$^-$, with one nuclear spin from $^{15}$N as quantum memory. The operation in the module only uses deterministic processes and interactions and achieves high fidelity ($>99$\%) operation, and modules are connected by optical fiber.  In the repeater node architecture, the processes between modules by photons can be in principle deterministic, however current limitations on optical components lead to the processes to be probabilistic but heralded.  The most resource modest repeater architecture contains at least two modules at each node, and the repeater nodes are than connected by telecom wavelength entangled photon pairs.  We discuss the performance of quantum repeaters starting from the minimum-resource strategy with several modules ($\sim 10$) and then incorporating more resource-intense strategies step by step.  Our architecture enables large-scale quantum information networks with existing technology.
\end{abstract}
\date{\today}

\maketitle

\section{Introduction}
The development of devices that process information according to the principles of quantum mechanics is leading to a new technological revolution. It is already clear that large scale quantum computers will be able to perform tasks impossible in the classical world,  however it is a daunting tasks to realise due to the huge number of physical qubits (billions at least) required \cite{DiV2009, Devitt2009, Ladd2010,Devitt2013}. The field of quantum communication is rapidly growing as it is seen as a simpler task than full large scale quantum computation \cite{Childress2005,VM2009}. Even though it is likely that resources of similar quality will be required, much fewer of them will be needed \cite{ Munro2010}.  

A key ingredient in any quantum communication network is quantum repeaters - devices that create entangled qubits between distance parties. The field has been working in two main directions: the first being the experimental realisation of small scale devices using high error rate components that unfortunately leads to poor communication performance\cite{Yuan2008} . The second has been theoretical work on large scale fully error corrected quantum systems, whose performance can be exceptionally fast \cite{Munro2012, Fowler2010, Li2013}.  However, small scale quantum computers required in such quantum communication systems are far from what can be realised with current technology.  One must bridge this gap to provide a viable route forward.

There are  many potential mechanisms to distribute entanglement remotely including ones based on emitters, receivers, transmitter and receivers. In this manuscript we illustrate how one can utilise the components and techniques being developed for large scale quantum computers in simple quantum repeaters - without initially having to resort to fully error corrected devices. Our approach is based on the state dependent reflection of an optical photon interacting  with an NV$^-$ center in diamond embedded in a cavity\cite{Nemoto2014}.  The scheme we present in this paper does not restrict its implementation to NV$^-$ centers in diamond.  However a number of good quantum properties and the required controllability with NV$^-$ centers have been demonstrated\cite{Maurer2012,Togan2010,Jelezko2004,Dutt2007,Neumann2008,Hanson2008,Jiang2009,Neumann2010,Neumann2010b,Buckley2010,Robledo2011,Sar2012,Dolde2013}, and it is therefore a promising candidate to implement such a scheme.  

The electron spin of a NV$^-$ center is used to mediate entanglement between optical photons (without direct excitation) and the nuclear spin-1/2 of $^{15}$N which is used as the long lived memory while optical signals propagate between nodes.  The same components can be used both for the entanglement distribution as well as the local two qubits gates. We however need to remember that NV$^-$ centers use an optical transition (637nm) but telecom wavelengths need to be used over optical fibers \cite{footnote2}. This necessities the use of frequency converters - to or from optical to telecom wavelengths. However moving from telecom to optical wavelengths is much easier than the optical to telecom process and so it is natural to consider a telecom based source of Bell states, rather than a single photon approach where both up and down conversion would need to be used \cite{footnote3}.

In the following we first describe the basic components of the repeater, what is needed to build a linear network and discuss its performance.  We then show how to boost performance by adding identical modules for multiplexing and error correction.

 \section{The module}\label{sec2}
The most important component for the repeater is a (quantum) data processing module.  We may additionally require optical elements such as photon detector, beam splitter, Bell state generators, single photon sources and coherent frequency converters.  The module is an interface between photon and matter qubits which store and process the quantum data.  We illustrate a design of such a module and its functions using a single negatively charged NV centre (NV$^-$) in an optical cavity.  The description given in this example can be applied to implement the same functions with other physical systems.  

The module consists of an optical cavity and a single NV centre (NV$^-$) in diamond \cite{Davies1976,Harley1984,VanOort1988} .  The single NV$^-$ centre provides an electron spin$-1$ and the nuclear spin$-1/2$ of $^{15}$N.  The Hamiltonian of the single NV$^-$ centre\cite{Robledo2011, Nemoto2014} is
 \begin{eqnarray}
 H_{nv} &=& \hbar ( D S_z^2 + E \left[S_x^2 - S_y^2\right]  + g_e \mu_B B S_z)\\
 & & -\hbar    g_n \mu_n B I_z\\
 & & +\hbar   A_{\parallel} S_z I_z+ \hbar \frac{A_{\perp}}{2} \left(S_+ I_- + S_- I_+ \right).
 \end{eqnarray}
%and the energy structure of the NV$^-$ is shown in the Fig.~\ref{energy}\cite{Robledo2011}.  
Here, the first term represents a zero field splitting ($D/2 \pi = 2.87$ GHz), a strain induced splitting ($E/2 \pi < 10\,$MHz), and a magnetic field induced splitting ($g_e \mu_B B$)  for the NV$^-$ centre's electron spin  \cite{Felton2009}. $S_{x,y,z}$ represents  the generalised Pauli $X$,$Y$,$Z$ operators for a spin-$1$ system with $S_+$ ($S_-$) being the raising (lower) operator.  The parameter $\mu_B$ is the Bohr magneton, and $g_e=2.0$ is the g-factor.  With an externally applied magnetic field of $B\sim 20$ mT, the $|0\rangle$ and $|+1\rangle$ levels at the ground manifold are separated by approximately $3.43$ GHz. The $|m_s=-1\rangle$ energy level is detuned approximately $1.1$ GHz below the  $|m_s=+1\rangle$ level and $\sim2.3$ GHz above the $|m_s=0\rangle$ level.  The electron states $|0\rangle$ and $|1\rangle$ in the ground state manifold span the Hilbert space of the electron spin qubit in the module.  
The second term represents a magnetic field induced splitting of the nuclear spin of $^{15}$N. $I_z$ is the Pauli $Z$ spin-$1/2$ operator, $\mu_n$ is the nuclear magneton, and $g_n=-0.566$ represents the nuclear g-factor. The computational basis states of the nuclear spin are  $|\downarrow\rangle$ ($|\uparrow\rangle$). 
The rest of the terms represents a hyperfine interaction between the electron and nuclear spins. The hyperfine interaction has both an Ising coupling with a coupling strength $A_{\parallel}$ and an exchange coupling with a coupling constant $A_{\perp}$.  For a $^{15}$N nucleus, $A_{\parallel}/2 \pi \sim 3.03$ MHz and $A_{\perp}/2 \pi \sim 3.65$ MHz \cite{Felton2009}.  

Now we turn to the cavity and the NV$^-$ electron spin.  We tune the cavity to be resonant to the energy gap between $|0\rangle$ states of the ground and the first excited states.  The electron spin states $|0\rangle$ and $|1\rangle$ are used to conditionally reflect incoming light field.  Assuming a high-cooperativity regime for the cavity; $C>>1$, the signal for a cavity with a NV$^-$ centre being in the ground $|0\rangle$ state is reflected as the reflection probability $P_r \sim 1$, while the signal for the empty cavity results in $P_r \sim 0$.\cite{Nemoto2014, Kimble1998}  In this regime, the electron state excitation decreases with cooperativity as $1/C$, however there is a small possibility that a off-resonant excitation of NV$^-$ centres in the $|1\rangle$ ground state may occur.  Most of times, an excitation is harmless, however due to a number of decay channels with non zero probabilities, it could cause a spin flip or an leakage error on the electron spin.  Such an error could transferred to nuclear spin through the hyperfine coupling.  Dealing with photons, we have to assume that it is inevitable for any gates mediated with photons to be probabilistic, and hence we need to design the gate to be heralded for success.  The heralding signal is given by the photon measurement, which guarantees that there was no excitation occurred in the case of success.  When the gate failed, we need to treat the electron and nuclear spins carefully.  If we can initialise both electron and nuclear spins after a failed event, it is straightforward to correct such errors.  We can treat the entanglement distribution between adjoint nodes in this way (details are discussed in Sec\ref{sec3}), however in general we have to repeat the gate sequence until success while the nuclear spin curries nontrivial quantum information, and hence such errors can be accumulated.  
To deal with such errors, it is shown that a use of an appropriately polarised optical field would be sufficient to suppress the unwanted excitations to meet the overall error rate for fault-tolerant quantum computation.\cite{Nemoto2014}.  

The electron spin coherence time so far is less than 0.1 ms, and the communication time for 10km through a fiber is approximately $50\mu s$, hence the electron spin coherent time is not long enough to maintain the quantum information with a high fidelity.  A coherence time for the nuclear spin of the $^{15}$N at the NV$-$ centre of 0.2s is expected, which is more than three orders of magnitude longer than the electron spin coherence time.  Hence the nuclear spin may be used as a quantum buffer or memory instead.  The hyperfine coupling can be used to realise a CZ gate between the electron spin and the nuclear spin.  The hyperfine interaction is always on, which could be a decoherence source for the nuclear spin, however we can effectively turn off the hyperfine coupling by setting the electron state to be $|0\rangle$.
   
 Now, we turn to single qubit rotations and measurements, which are essential for initialisation, single-qubit operation, and readout.  We start with spin rotations.  The spin rotations can be implemented via an electromagnetic driving field, the interaction can be given as
 \begin{equation}
  H_d=\hbar\Omega_0 \cos (\omega_d t+\phi)(S_x-\frac{g_n\mu_n}{g_e\mu_B}I_x),
  \end{equation}
where $\Omega_0$ is the amplitude of the applied field.  The frequency $\omega_d$ is chosen to determine whether we drive the electron or nuclear spin with the specific phase $\phi$.  
The initialisation and readout can be done through measurements.  The projective measurement of the electron spin on to the computational basis can be implemented as QND measurement via photon.  The conditional reflection introduced previously can be used to determine whether or not the electron spin state is $|0\rangle$ or $|1\rangle$ by detecting the reflected photons.  Combining it with single rotations, we can implement $X-$ and $Y-$ measurements.  Such QND measurements can be also done with weak coherent state, however to achieve a higher overall efficiency, we use repeated QND measurement with single photons.  This way we avoid deionisation of NV$^-$ as well as minimise unwanted excitations to the first excitation manifold\cite{Tamarat2008,Siyushev2013}.  There is a small possibility that the electron spin state leaks to $|-1\rangle$.  Through the repeated QND measurement, we can detect such leakage and reset the electron spin by the spin rotation.  The projective measurement on the nuclear spin can implemented via the hyperfine interaction and the QND measurement on the electron spin.\cite{Nemoto2014}  The natural hyperfine interaction enables fast $Z-$basis measurements, while $X-$ and $Y-$ basis measurements can be done with a driven hyperfine interaction.\cite{Everitt2014}.  Combining the measurement schemes and the single qubit rotations, we can initialise the state of both electron and nuclear spins.
 
As we described above, the NV$^-$ centre in an optical cavity is a good candidate to construct the module, however this is not only the implementation possible.  As long as all the module functions are satisfied with required fidelities, a design radically different from this may be considered. The specific physical parameters are necessary for the construction and evaluation of the module and its systems \cite{Nemoto2014}.

 \section{Remote Entanglement Distribution}\label{sec3}
\begin{figure}[htb]
\includegraphics[scale=0.45]{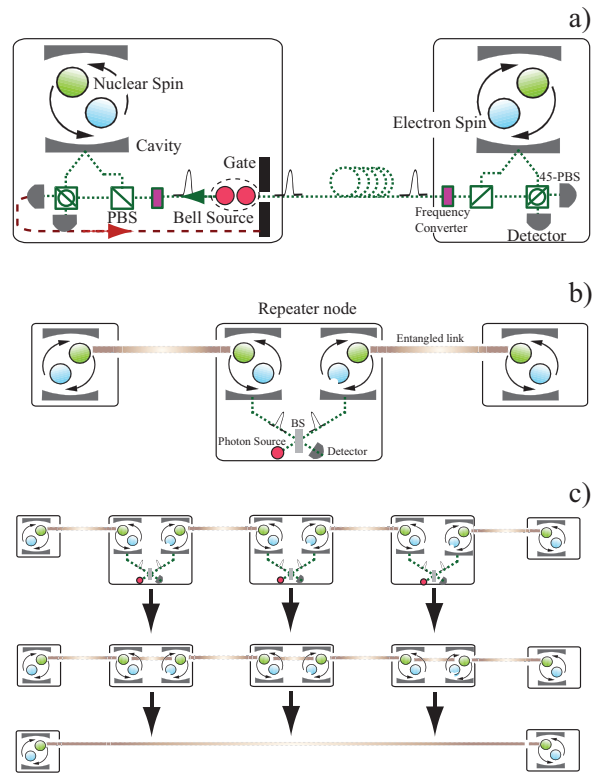}
\caption{a) Schematic diagram of an entanglement generation between two remote nodes connected by fiber.  The nodes are composed of a NV$^-$ center embed in a single sided optical cavity along with a frequency converter from the telecom wavelength photons to the NV center transition and detectors that measure in the diagonal basis. The left hand side nodes also includes a source of $\frac{1}{\sqrt 2}\left[ |HV\rangle-|VH\rangle\right]$ entangled photons  at the telecom wavelength and gating operations. The gate opens only upon a successful measurement event at this node. b) Entanglement swapping within a repeater node based on the probabilistic entanglement of the individual NV$^-$  electron spin using a single photon in a Mach Zender interferometer arrangement. Detection at the dark port implies a maximally entangled Bell state has been generated. Using the hyperfine interaction and electron spin measurements, the entanglement link can be transferred to the nuclear spins - effectively creating a chain of entangled nuclear spins \cite{Nemoto2014}. c) Entanglement swapping over many repeater nodes.}
\label{entanglementdist}
 \end{figure}

Our scheme for the remote entanglement distribution is depicted in Fig. (\ref{entanglementdist}a).  This scheme is applied to establish entanglement between two adjoint nodes.  As we mentioned in Sec.\ref{sec2}, the cavity of the module is tuned to conditionally reflect an incoming photon only when the electron spin state is $|0\rangle$, and we use this conditional reflection to establish entanglement between two electron spins in different modules.  As shown in Fig. (\ref{entanglementdist}a), a high-rate Bell source as well as a polarisation selective detector is inserted at the sender node.  The creation of the link begins with the Bell source emitting an entangled pair, $\frac{1}{\sqrt 2}\left[ |HV\rangle-|VH\rangle\right]$, at the telecom wavelength. One photon, to be sent to the receiver node, is temporally buffered while the second is frequency converted so it can interact at near resonance with the cavity containing an NV$^-$ centre electron spin prepared in the state $|+\rangle= \frac{1}{\sqrt 2} \left[ |0\rangle+ |+1\rangle \right]$.  Then, a $\pi$ phase shift is applied to the vertically polarised photon only when the qubit is in the $|0\rangle$ state. The photon is then measured in the diagonal (D) basis giving  a result $D$, $A$ and $0$ where $0$ no photon detected.  

%{\bf Bill: put more explanation}

In the event of a photon detection, the electron spin in the sender module is entangled with the fist photon stored in the buffer.  Before sending the first photon, we store the shared entanglement with the photon to the nuclear spin, of which coherence time is significantly longer than the electron coherence time, avoiding the deterioration of the fidelity from the electron spin dephasing.  This entanglement transfer operation can be done via the hyperfine coupling and the electron spin measurement and initialisation.  Then, the buffered photon is transmitted over the link (for example in time bin encoded format) to the adjacent nodes along with the measured detection result (either D or A).  Upon arrival of the photon at the remote receiving node, its frequency is converted to the optical wavelengths where it then interacts with the second NV$^-$ center also prepared in the $|+\rangle$ state. Again a $\pi$ phase shift occurs on the vertically polarised photon if the electron spin was in the $|0\rangle$ state. The polarised photon is then measured in the diagonal basis.  When the photon is successfully measured, the entanglement is stored in the nuclear spin again via the hyperfine coupling and the electron spin state measurement.  The resulting state, dependent on a successful measurement result, is 

\begin{eqnarray}
\rho(t)&=&\frac{ 1+e^{- t / T_2}}{2} |\zeta_-\rangle \langle \zeta_-| + \frac{ 1-e^{-t / T_2}}{2} |\zeta_+\rangle \langle \zeta_+|,  \nonumber \\
\end{eqnarray}
giving a resulting Bell pair is $ |\zeta_-\rangle =  |\Phi_-\rangle,  |\zeta_+\rangle= |\Phi_+\rangle$ for the $D,D$ and $A,A$ events and $ |\zeta_-\rangle =  |\Psi_-\rangle,  |\zeta_+\rangle= |\Psi_+\rangle$ for the $D,A$ and $A,D$ events with $|\Phi_\pm\rangle = \frac{1}{\sqrt 2} \left[|0\rangle |0\rangle \pm  |+1\rangle |+1\rangle\right]$ and  $ |\Psi_\pm\rangle = \frac{1}{\sqrt 2} \left[ |0\rangle |+1\rangle \pm |+1\rangle |0\rangle\right]$ with the fidelity $F(t)=\frac{ 1+e^{- t / T_2}}{2}$ . Here $T_2$ is the coherence time of the nuclear spins while $T_R$ ($T_R \ll T_2$) is the roundtrip time for a signal to propagate between adjacent nodes \cite{footnote4}. Each of these events (D,D, D,A, A,D, A,A) occurs with a probability 
\begin{equation}
e^{-L/L_0}  (1-T) p_c p_D/8, 
\end{equation}
thus giving an overall success probability  
\begin{equation}
p_s = e^{-L/L_0}  (1-T) p_c p_D/2.
\end{equation}
Here $L$ is the length of the channel between the nodes, $L_0$ the attenuation length of the fiber, $p_D$ the single photon detection probability, $p_c$ the coupling efficiencies associated with the cavity interaction (including the frequency conversion frequency) and $T$ the transmission coefficient ($T\sim -1$) \cite{footnote5}. If the detection event was $0$, the procedure has failed and the protocol needs to start again.  At the stage of the entanglement generation, there is no communication information involved in the procedure, and hence the electron and the nuclear spins can be measured and reinitialised for the next round of entanglement generation without any loss of communication information.  

\subsection{Performance and Rates}
Each attempt of the entanglement generation described above is probabilistic in nature, and this basic entanglement link can be converted to near deterministic with the use of spacial or temporal resources.  With limited initial physical resources, the temporal approach (a repeat until success strategy) should be used scarifying the operational time.  To achieve a failure probability $\epsilon$ for the basic entanglement link, $n \sim \log \epsilon / p_s$ attempts are required.  After the $n$ attempt, the missing link probability $\epsilon$ reduces the rate of the basic entanglement link, hence the rate is
\begin{eqnarray}
R &\sim&  - (1-\epsilon) \frac{ c}{2 L} \frac{p_s}{ \log \epsilon} \\
&\sim&  -  (1-\epsilon)\frac{ e^{-L/L_0}}{L} \frac{c\; p_c p_D (1-T) }{ 4  \log \epsilon}.
\end{eqnarray}
We could have the rate $R$ to be $\frac{ c}{2 L} \frac{p_s}{ \log \epsilon}$ incorporating the failure factor $\epsilon$ to the fidelity of the Bell pair.  However, with the healed signal, we know when the link failed, and hence we can keep the fidelity untouched scarifying the generation rate.
 
\section{Simple Linear Chains}
 
 The next step is to move from two adjoint nodes to a linear chain of quantum repeater nodes. Among a number of approaches and strategies that can be used to implement a linear chain quantum repeater, the simplest is the minimal resource approach by Lukin et. al\cite{Childress2005}, where each node has one NV$^-$ center hosting two qubits (an electron spin and a nuclear spin).  However the electron spins coherence time significantly limits the quality of the longer range entangled links that can be constructed. Instead by moving to two NV$^-$ centers per node (Figure \ref{entanglementdist}b), one can use the nuclear spins long coherence properties to establish all the long range link, yet use the electron spins for interface to distribute entanglement both within and between the nodes. The local two qubit gates between NV centers in different cavity within the same node are mediated via an optical link which facilitates the two electron spins to be entangled.  This electron spins entanglement can then be transferred to the nuclear spins creating an entanglement chain, i.e. a linear cluster state, of NV center electron and nuclear spins \cite{Nemoto2014}.  By measuring the nuclear spins in the intermediate repeater nodes in an $X$ basis, they are disentangled and a longer range nuclear spin entangled link created (Figure \ref{entanglementdist}c).

\begin{figure}[hbt]
\includegraphics[scale=0.9]{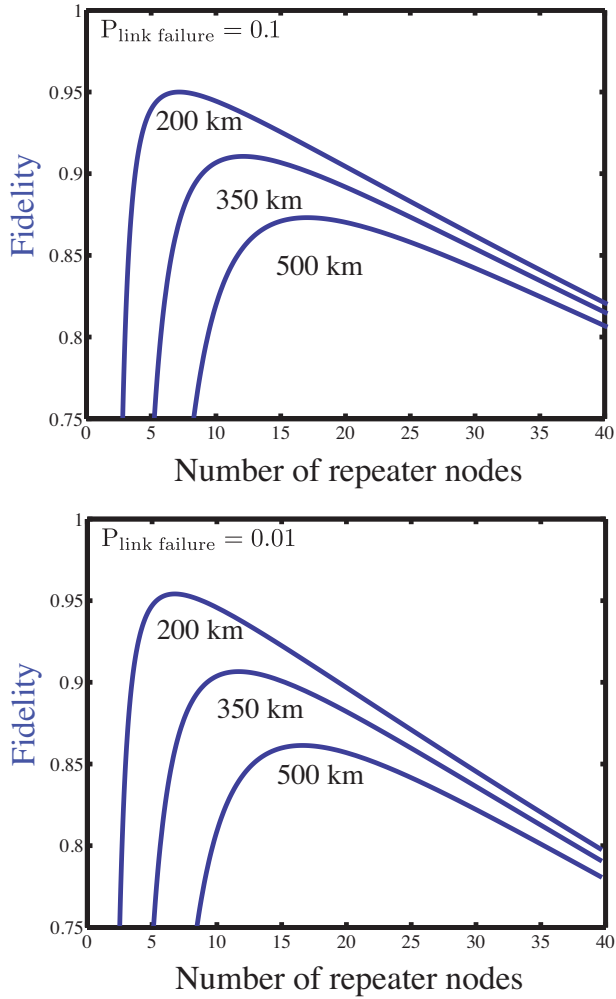}
\caption{Fidelity of the resulting Bell state between the end node of the repeater chain versus $N$ for for network distances of 200, 350 \& 500 km. }
\label{performancefid}
 \end{figure}

Given the high initial fidelity of the nuclear spin links and the near deterministic local gates, we can now estimate the performance of such a repeater scheme. One can also easily establish an estimate for rate to generate Bell pairs between the end nodes in the linear chain as
\begin{eqnarray}
R_{\rm net} \sim    -  (1-\epsilon)^N \frac{N e^{-L_{\rm tot} / N L_0}}{L_{\rm tot} } \frac{c\; p_c p_D (1-T) }{4\log \epsilon}
\end{eqnarray}
where $N+1$ is the number of repeater nodes in the chain. The expression of $R_{\rm net}$ indicates that increasing the number of repeaters nodes (increasing N) for the fixed distance $L_{tot}$ would give a better performance, however as our initial entangled links have a finite fidelity $F$, performing $N-1$ entanglement swapping operation would lead to a decrease in the overall fidelity $F_{\rm net} = F^N F_{gate}^{N-1}$, where $F_{gate}$ is the gate fidelity for performing the swapping operation. As the number of nodes enhances the degradation of the fidelity, with more nodes in the repeater chain we end up with the worse the fidelity of the resulting entangled link between the end nodes. To illustrate this we plot the resulting fidelity in Figure (\ref{performancefid}) versus $N$ for a 200 km, 350km \& 500 km network. 
The high initial fidelity links between adjacent repeater nodes plus fast, near deterministic and efficient local gates indicate that purification is not required as long as the number of nodes is small. For the smaller number of nodes, the probability of success for a single link generation between adjacent nodes is quite low and so many attempts are needed ($>100$).  Although it requires less entanglement swapping gates, generating an end to end eventually approach to the life time of the nuclear spin and hence its coherence time limits our Bell pair fidelity, which appears in the sharp raising curve in the small node-number domain in Fig. \ref{performancefid}. For a much larger $N$, the fidelity is instead limited by the gate fidelity $F_{gate}$, shown in the large node-number domain in Fig. \ref{performancefid}.

An issue arises where one wants to make comparisons with different numbers of repeater nodes. Changing the number of nodes or the number of qubits within a node will dramatically change its rate of communication. Hence some form of resource normalisation could be appropriate. There are many ways this could be achieved but a natural one would be to divide the rate by the total number so qubits in the whole network. In such a case this normalised rate $\bar{R}_{\rm net}$ can be estimated as
\begin{eqnarray}
\bar{R}_{\rm net} \sim    - (1-\epsilon)^N \frac{e^{-L_{\rm tot} / N L_0}}{L_{\rm tot} } \frac{c\; p_c p_D (1-T) }{8\log \epsilon}.
\end{eqnarray}
%where we immediately notice that to first order that it does not depend on $N$. 

% For an initial bell pair with fidelity $F$ between adjacent nodes and local gates with fidelity $P_{\rm local\;gate}$, then an $N+1$ node network has a fidelity of $F_{\rm link}^N F_{gate}^{N-1}$. We plot the resulting fidelity in Figure (\ref{performancefid}) versus $N$ for a 200 km, 350km \& 500 km network. The high initial  fidelity links between adjacent repeater nodes plus fast, near deterministic and efficient local gates means purification is not required as long as the number of nodes is small. For the smaller number of nodes, the probability of success for an initial try of a single link between adjacent nodes is quite low and so many attempts are needed ($>100$). However this means our time for generating an end to end start to approach the life time of the nuclear spin and so its coherence time limits our Bell pair fidelity. For much larger $N$ we become limited by the gate fidelity $F_{gate}$.

\begin{figure*}[hbt]
\includegraphics[scale=0.9]{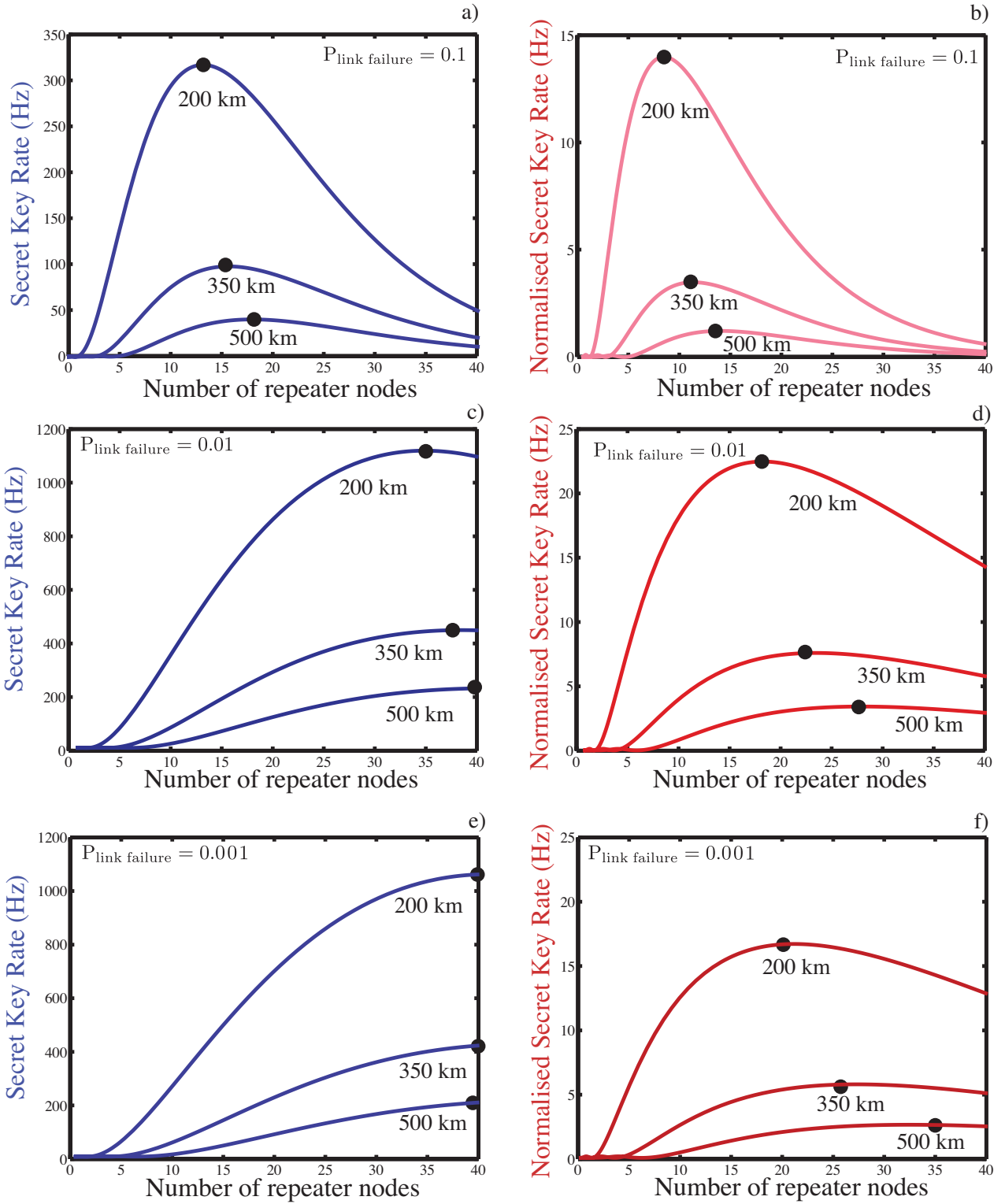}
\caption{ Long distance secret key rates (a,c,e) and normalised secret key rates (b,d,f) for an $N+1$ node linear repeater network  with total distances of 200, 350 \& 500 km for various link failure  probabilities $p_{\rm link\;failure}=\epsilon=0.1,0.01,0.001$. The network has $N-1$ intermediate repeater nodes with each node containing two NV$^-$ centers. The secure key rate is calculated by multiplying the end to end link rate by $1-H \left[\epsilon_f\right]$ where $H[x]$ is the binary entropy function and $\epsilon_f$ the error rate of the end to end Bell pair. The normalised secure key rate is calculated by multiplying the end to end link rate by $1-H \left[\epsilon_f\right]$ divided by the total number of NV$^-$ centers used in the  network. In c) with $p_{\rm link\;failure}=0.01$, the optimal separation between the repeaters nodes are 5.71 km, 9.21 km and 12.20 km respectively, while for d) it is 11.11 km, 15.21 km and 17.86 km respectively.   }
\label{performance2}
 \end{figure*}

\subsection{Secret key rate}

Being able to determine the rate and fidelity of entangled state generated over the end points of the networks gives us very useful information about the performance of our scheme. As mentioned earlier one of the natural applications for a long range entangled link is quantum key distribution. In most QKD scheme the keys can only be established over a maximum distance of 200 km and at this extreme range, the rate is generally quite low\cite{QKD,QKD2}. Given our repeater scheme, we can now determine the secret key rate and normalised secret key rate by the number of devices one can generate over 200, 350 and 500 km. Our secret key rate $C_r$ (normalised secret key rate $\bar{C}_r$ ) is given by the rate $R_{\rm net}$ (normalised rate $\bar{R}_{\rm net}$) of generating entangled links multiplied by the factor of secret keys material per Bell state. This can be expressed as 
\begin{eqnarray}
C_r = R_{\rm net} &\times& (1-H \left[\epsilon_f\right]),\nonumber \\
\bar{C}_r = \bar{R}_{\rm net} &\times& (1-H \left[\epsilon_f\right])
\end{eqnarray}
respectively where $H[x]=-x \log_2 x - (1-x) \log_2 (1-x)$ is the binary entropy function and $\epsilon_f= 1- F_{\rm net}$ the error rate of the end to end Bell pair.  This is due to the dominant phase error in the system, this simple entropy function is enough for our estimation.  In Figure (\ref{performance2}) we plot the secret key and normalised secret key rate versus the number repeater links for three choices of the failure probability for a link being generated between adjacent nodes, $\epsilon=0.1,0.01,0.001$.  Figure (\ref{performance2}) shows the different performance for each $\epsilon$.  For $\epsilon=0.1$ there is 10 percent chance an individual link will failure and in this case the overall chain failures.   This means one does not want too many repeater nodes present (which can be seen from the peak maximum between 10 and 20). However for $\epsilon=0.001$ the failure probability is on the order of 0.1 percent and so for $N<100$ links do not fail very often. This means it should be rare for the entanglement chain to be broken but the cost will be we wait a long time for the individual nodes between adjacent repeater nodes to be generated and so the overall rate  could be low. 

An $\epsilon$ value between these should have better performance which can be seen in Figure (\ref{performance2}c,d). What is quite interesting is the optimal number of repeater nodes  depends heavily on whether one is considering the raw or normalised rates. As we add extra nodes to the repeater chain, the raw rate can obviously increase until the loss in fidelity balances it out. However for the normalised rate, we also need to divide by the $2 N$ NV$^-$ centers used in the linear network and so we would expect the optimal point to be reached for fewer repeater nodes. Thus we expect the number of nodes to be less in this second case. 

The low rate key rate for small numbers of repeater nodes is again due to the probability of success for each try of a single link between adjacent nodes being quite low, and hence many attempts are needed. This means the time to generate our end to end entangled links starts to approach the life time of the nuclear spin and so the fidelity of the resulting state is low. For large $N$ we fidelity we also limited by the $F^N F_{gate}^{N-1}$ fidelity.  This two effects compete with one another and so there is an optimal point for moderate $N$.  Having a repeater scaling between 10 - 20 km apart seems to be a good working point. 
%%JS: for which nuclear spin lifetime are the plots. need to specify.

\section{Multiplexing}
 
\begin{figure}[htb]
\includegraphics[scale=0.75]{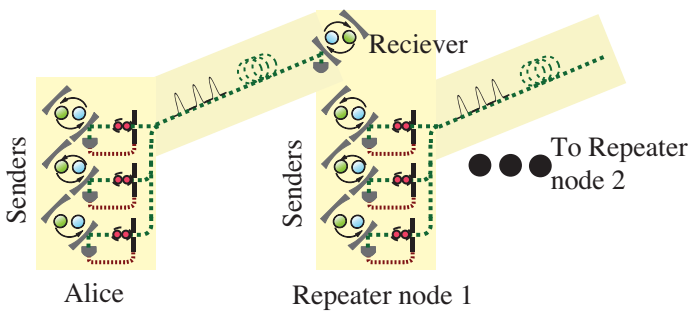}
\caption{Schematic diagram of a spatially multiplexed quantum repeater scheme. }
\label{fusiliers}
\end{figure}

Our performance for creating longer range links is primarily limited  by the time to create the entangled links between adjacent nodes, and not by the local gates to perform the swapping operations. When the nodes separation exceeds the attenuation length of the fiber, we need to wait a significant number of round trip time for quantum/classical to be sent between adjacent nodes. This is a form of temporal multiplexing but its also has a secondary detreminal effect on our NV$^-$ centers due to their finite coherence times. This memory issue can be overcome by using a spatial multiplexing strategy costing more physical resources. However more efficient spatial strategies can increase the performance  further \cite{Munro2010} as we depict in Fig (\ref{fusiliers}) as it uses approximately half the resources of the previous schemes. 
With $n$ senders and 1 receiver, the probability a link is established is $p_S=1-\left(1-p_s\right)^n$.  Of course more than one link between adjacent repeater nodes can be established at the same time. In fact, if one requires q copies then the success probability for n senders is $p_S= 1- \sum_{i= 0}^{q-1} \left( n\atop i \right)  p_s^i \left(1-p_s\right)^{n-i}$. These $q$ copies can be used in a number of ways including:
\begin{itemize}
\item Increasing the rate for generating long range Bell states. 
\item Performing some form of error correction to increase the range and fidelity of the long range Bell states. 
\end{itemize}
 
\begin{figure}[hbt]
\includegraphics[scale=0.9]{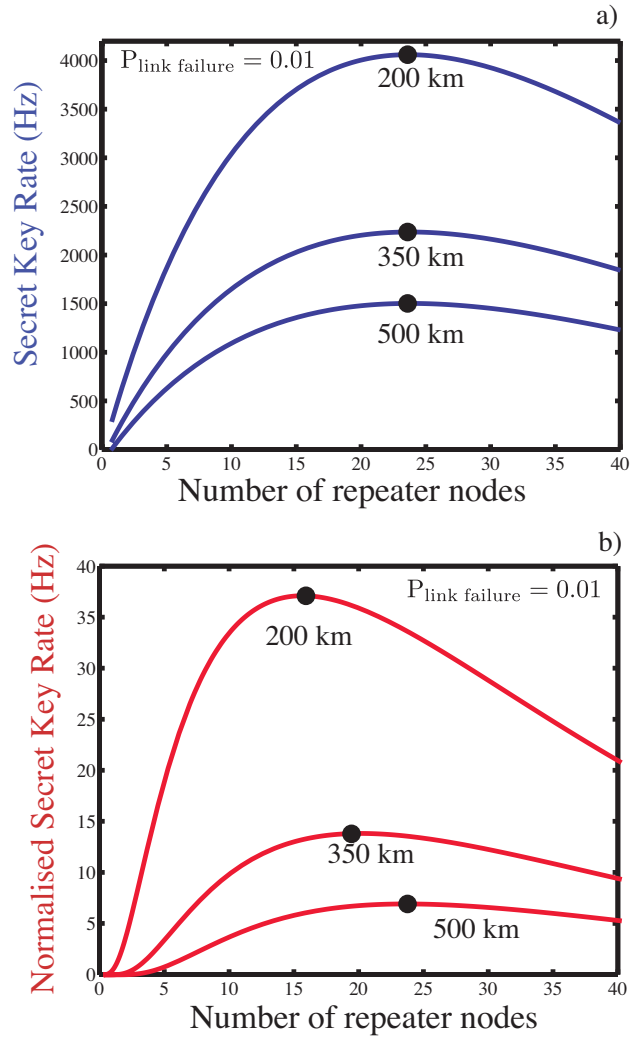}
\caption{Long distance secret key rate  a) and normalised secret key rate b) for an $N+1$ node linear repeater network  with total distances of 200, 350 \& 500 km for a link failure  probabilities $p_{\rm link\;failure}=\epsilon=0.01$. The network has $N-1$ intermediate repeater nodes with each node containing $m+1$ NV$^-$ centers.  In a) the optimal separation between the repeaters nodes are 8.33 km, 14.58 km and 20.83 km respectively for the 200, 350 \& 500km overall distances, while for b) it is 12.5 km, 16.67 km and 20.83 km respectively.}
\label{performancemult}
 \end{figure}

Let us consider the first item.   In Figure (\ref{performancemult}) we show the long distance secret key rate and normalised secret key rate for several network lengths, using the multiplexed strategy depicted in Fig (\ref{fusiliers}). Within a repeater node we have $n+1$ qubits, $n$ qubits used to establish entanglement to it right hand neighbour and one to acts as a receiver to accept connections coming from the left hand adjacent repeater node. Of course as we are using more qubit per node, our raw (un normalised) secret key rate increase. The normalised rate however also increases and this is primarily due to removing the detrimental memory effects.

 \begin{figure}[hbt]
\includegraphics[scale=0.9]{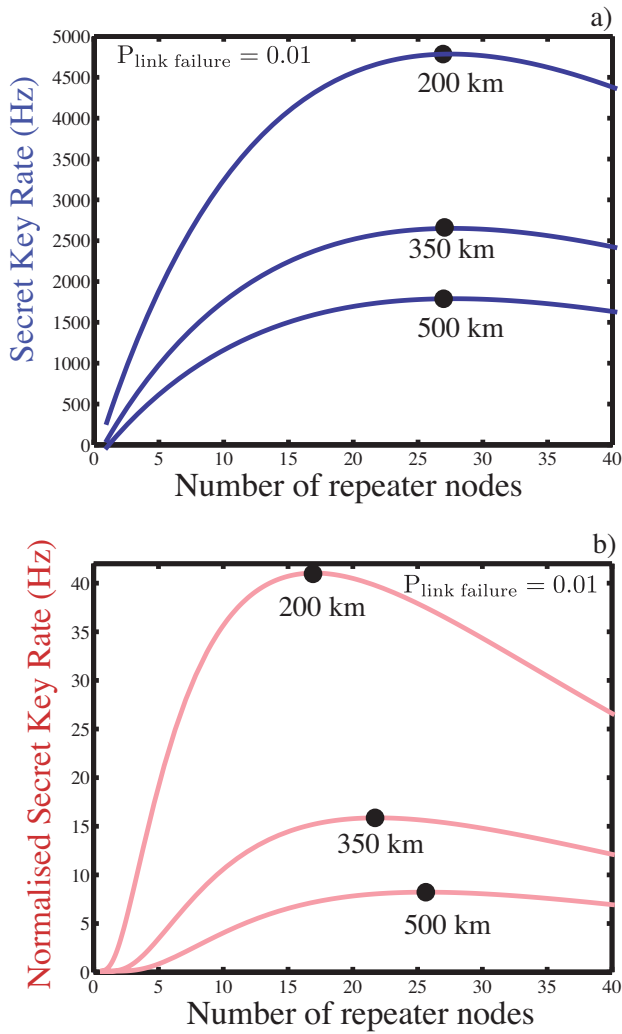}
\caption{ Long distance secret key rate  a) and normalised secret key rate b) for an $N+1$ node linear repeater network  with total distances of 200, 350 \& 500 km for a link failure  probabilities $p_{\rm link\;failure}=\epsilon=0.01$ when more than 1 qubit of information is encoded on the photon. The network has $N-1$ intermediate repeater nodes with each node containing $m+q$ NV$^-$ centers.  In a) the optimal separation between the repeaters nodes are 7.41 km, 12.5 km and 17.86 km respectively for the 200, 350 \& 500km overall distances, while for b) it is 11.76 km, 15.91 km and 19.23 km respectively.}
\label{performance}
 \end{figure}
 
The second case is quite interesting as it opens another possibility for how we send information between the nodes. We could now encode multiple bits of information (say q bits) onto the transmitted single photon using time bin, frequencies etc \cite{footnote7}.  In such a case we create a Bell state composed of the NV$^-$  center (one qubit) and a hyper-encoded photonic state (encoded qubit). This hyper-encoded photonic state is transmitted over the network to the receiver side where its state is transferred back to $d$ qubits and decoded to correct errors, that had occurred during the transmission or on the receiver side (It however will not correct memory based errors on the original bare  NV$^-$  center qubit).  The hyper encoding allows us to improve the key generation rate as is shown in the red curves in Fig (\ref{performance}). However such a strategy does not allow us to significantly  increase the total distance entanglement can be generated due to the imperfect Bell pairs created between adjacent nodes and errors associated with the local gates ( $\sim 0.3\%$). It is difficult to have a normalised key rate greater than 1 bit/s for distance greater than 1000 km. 

 \section{Going longer : Error correction}
 
\begin{figure}[hbt]
\includegraphics[scale=0.45]{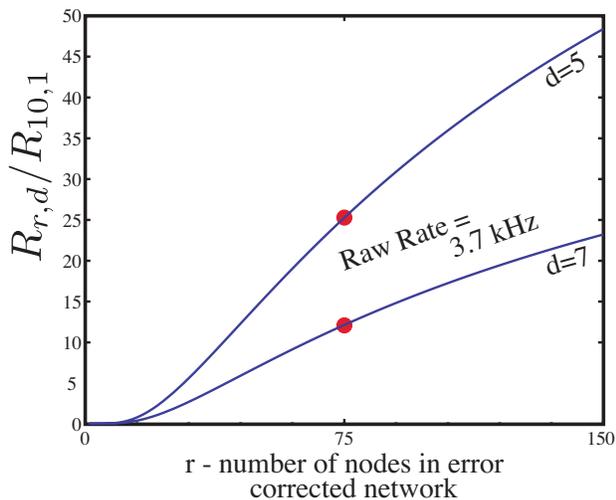}
\caption{Normalised rate $R_{r,n_d}$ (for a 2000 km fully error corrected network (with codes distances of d=5 and d=7) compared with the rate $R_{10,1}$ of a 10 link repeater network without error correction. The raw communications are is 3.7 kHz for both the d=5 and 7 codes. Both achieve a final fidelity for their entangled pair grater than 99\% (99.9\% and 99.99\% respectively). The fidelity of the straight (no error correction) 10 link network is approximately 92\%.}
\label{improvementregion}
 \end{figure}

 To establish longer links one needs to perform either long range purification or error correction. Usual pair-wise  purification is not ideal as it requires extensive classical communication which significantly limits its performance and the extensive classical communication dramatically increases the requirements on the quantum memories\cite{Mohsen2013}. Error correction could exhibit similar limitations, however error correction provide different ways to protect coherence of the state\cite{Stephens2013}, and hence its use does raise a number of important issues. First and foremost is the effect on performance by doing the error correction itself. To perform error correction we need many entangled links between adjacent nodes and so one would think that the rate of communication would decrease. For a distance $d$ error correction code, $n_d$ entangled pairs are needed.  It is straightforward to show the rate for generating an encoded entangled link over a distance of $L_{\rm total}= N L$  divided by the total number of qubits used in the network is 
\begin{eqnarray}
\bar{R}_{\rm net}(N,n_d) \sim    -  \frac{e^{-L_{\rm tot}/N L_0} }{n_d L_{\rm tot}} \frac{c\; p_c p_D (1-T) }{4 \log \epsilon}
\end{eqnarray} 
We immediately notice that the normalised rate is lower by a factor of $n_d$ compared to case if not error correction had been done.  However this does not indicate that error correction does not give us any improvements on the normalised rate.  To show this, we could ask if any strategies with error correction can beat a non-coded strategy.  To show this quantitatively, we assume a 10 link linear quantum repeater over 2000km, which gives the normalised rate $\bar{R}_{\rm net}(10,1)$, and see if there is any strategy with error correction of the normalized rate $\bar{R}_{\rm net}(r, n_d)$ that can exceed $\bar{R}_{\rm net}(10,1)$.  In Figure (\ref{improvementregion}) we compare the performance using normalised rates of two error corrected codes with distances of d=5 and 7, which are based on topological codes, requiring $n_q=$81 and 169 respectively qubits per node in this scenario.  The plots clearly show that once the number of nodes is above 20, a rate improvement can be obtained. More specifically for 75 nodes, the improvement is 45 times for the d=5 code and 14 times for the d=7 code. The d=7 code however does give a much higher fidelity pair than the d=5 case (by at least one order of magnitude).  This does lead to a natural question of what the requirements are for an improvement in rate. 

 \subsection{Improvement criteria}
\begin{figure}[hbt]
\includegraphics[scale=0.45]{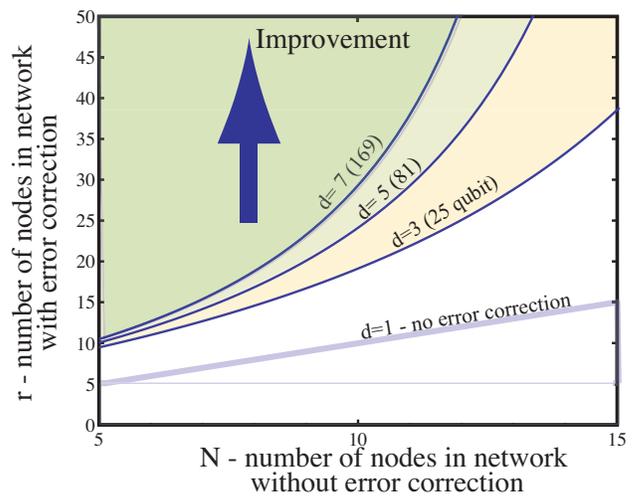}
\caption{Comparison of the rate of generating an entangled link over a  2000 km network using various topological error correction codes of distance $d=3,5,7$ (with qubit numbers 25,81,169 respectively).   Here $N+1$ is the number of rhodes in the case of no error correction while $r$ is the number of nodes in the error corrected case. Improvement in the normalised rates are shown as shaded areas above the corresponding $d$ boundary. }
\label{improvementregiona}
 \end{figure}

Now we formulate improvement possibility by error correction as an improvement criteria. as the range of $r$ to satisfy $\bar{R}_{\rm net}(r,n_d) >\bar{R}_{\rm net}(N,1)$.  This gives the condition for $r$ as
\begin{eqnarray}
r> \frac{N L_{\rm tot}}{L_{\rm tot}- N L_0 \log_e n_d},
\end{eqnarray}
An improvement can only be obtained when $L_{\rm tot}> N L_0 \log_e n_d$ and $r>N$.  Further if the separation between nodes $L = L_{\rm tot}/ N$ is less the attenuation length of the fiber $L_0$, error correction can not increase the performance. To illustrate this we show in  figure (\ref{improvementregiona}) the boundaries for different $n_d$ with $L_{\rm tot}=2000$ km where $R_{r,n_d} > R_{N,1}$. Above these boundaries an overall improvement to the normalised rate occurs. 
%For instance with $d=25$ we can almost get an improvement of nearly two orders of magnitude in rate. 

%[4/9/14, 3:25:40 PM] Ashley Stephens: distance 3, 3.1e-05
%[4/9/14, 3:25:47 PM] Ashley Stephens: distance 5, 2.4e-07
%[4/9/14, 3:26:01 PM] Ashley Stephens: distance 7 will take longer because the failure rate is so low
%
%
%distance 3, 25 qubits, 1.4e-3
%distance 5, 81 qubits, 1.5e-4
%distance 7, 169 qubits, 1.3e-5
%distance 9, 289 qubits, 9.1e-7
%distance 11, 441 qubits, 6.0e-8
%distance 13, 625 qubits, 4.0e-9

\section{Conclusion}
To summarise, we have presented a simple repeater scheme based on NV centers in diamond which can be used in a few node network yet scaled to a large scale networks as more resources become available.  For shorter distances, 200, 350, 500km, a simple scheme of two NV centers for each node can give significant gain in its secret key rate.  Although the setting is not optimal, a repeater with ten NV-enters can exhibit a significant gain over 200km.  For a longer distance, The performance shown in Fig. \ref{performance2} explains the tradeoff between the longer waiting time for entanglement distribution in two adjoint nodes and the increasing swap operations with a larger node number.  This simple scheme does not have a mechanism to improve the waiting time nor to recover from the increasing gate errors, we can address these issues by introducing multiplexing for the former and error correction for the latter.  The improvement multiplexing is shown in Fig. \ref{performance}.
  The improvement error correction can give has a richer properties.  As error correction imposes a significant constance overhead in the resource, i.e. the number of NV centers, an interesting question is error correction could ever give us any benefit for normalised Bell-pair generation rate.  Error correction pushes the fidelity of the final Bell-pair, hence if we have improvement in the net rate, a better improvement is guaranteed for its secure key rate.  It could be interesting to see how the secure key rate behaves, however as error correction can keep the fidelity very high in comparison to non-error corrected schemes, secure key rates do not fully capture the properties of quantum repeaters.  For instance, error correction allows to achieve 99.9\% final fidelity, enough for fault tolerant quantum computation and communication, which requires roughly an error rate of the order of $10^{-5}$ for local gates for a 20 link linear repeater. 
 To conclude, we evaluated the performance for the simple, two-NV centre per node scheme, the multiplexed scheme, and the error corrected schemes.  This indicates that with such devices as a NV-center based cavity device, can realise a quantum communication system which clearly show the figure of merit of quantum repeater.  The performance also shows how we can extend a simple scheme can be extended towards fully fault tolerant quantum communication.  Although the performance has been evaluated in linear network setting, the device allows as to implement arbitrary shaped network by classical switching.


\begin{thebibliography}{99}
\bibitem{footnote2} Up-conversion from telecom to optical wavelength is high efficient and near deterministic. However downconversion from optical to telecom is highly inefficient and probabilistic.
\bibitem{footnote3} Over long distances it is also extremely challenging to stabilise the larger interferometers needed for such protocols.
\bibitem{footnote4} In this situation we assume the central node in Figure (\ref{fusiliers}) perform entanglement swapping as soon as the links to the left and right are established. No waiting time is required for the memories in this case assuming one is careful in arranging how the links are connected. However the left and right outer nodes have to wait on the roundtrip time $T_R$ before they can use their resources. Hence one roundtrip time of dephasing on the nuclear spins arises.
\bibitem{footnote6} The no detection event can occur if the photon has been lost in the circuit or if the entangled pairs was not created successfully
\bibitem{footnote7} There is no advantage in sending multiple qubits on one photon to establish independent links as this linearly uses more resources on the repeater nodes and so does not improve the resource normalised performance.
\end{thebibliography}
\end{document}